\documentclass[preprint]{revtex4-1} 

\usepackage{graphicx}
\usepackage{natbib}
\usepackage{color}

\begin{document}

\title{Three-wave interaction and Manley-Rowe relations in quantum hydrodynamics}

\author{Erik Wallin}
\affiliation{Department of Physics, Ume{\aa } University, SE--901 87 Ume{\aa}, Sweden}

\author{Jens Zamanian}
\affiliation{Department of Physics, Ume{\aa } University, SE--901 87 Ume{\aa}, Sweden}

\author{Gert Brodin}
\affiliation{Department of Physics, Ume{\aa } University, SE--901 87 Ume{\aa}, Sweden}

\begin{abstract}
The theory for nonlinear three-wave interaction in magnetized plasmas is
reconsidered using quantum hydrodynamics. The general coupling coefficients
are calculated for a generalized Bohm de Broglie term. It is found that the
Manley-Rowe relations are fulfilled only if the form of the particle
dispersive term coincides with the standard expression. The implications of
our results are discussed.
\end{abstract}


\maketitle

\section{Introduction}
During the last decade much work has been devoted to quantum plasmas, see
e.g. Refs. \citep{Haas-book,Manfredi,Shukla-Eliasson,Shukla-Eliasson-RMP,glenzer-redmer} and
references therein. Laboratory applications include quantum wells \citep%
{Manfredi-quantum-well}, spintronics \citep{Spintronics} and plasmonics \citep%
{Atwater-Plasmonics}. Quantum plasma effects can also be of interest in
experiments with solid density targets \citep{glenzer-redmer}, as well as in
astrophysics \citep{Astrophysics,Astrophysics1,Astrophysics2}.

Nonlinear wave-wave interaction in plasmas has been studied since the
sixties, see e.g. Refs. \citep{Sagdeev-64,Sjolund-67,Kadomtsev,Tsytovich}. Of
special interest here is the three wave interaction processes, that have a
wide range of applications, including e.g. stimulated Brillouin scattering
in the ionoshere \citep{Dysthe-1977,Stenflo-2004} and various processes in
laser-plasma experiments \citep{Lashmore-Davies-book-chapter,Kruer-book-laser-plasma,Mironov-1990}. From a
theoretical point of view the Manley-Rowe relations \citep{Manley-Rowe} are
of much interest when three-wave processes are studied \citep{Weiland-Wilhelmsson,Larsson-1973,Larsson-1977,Brodin-88}. For example, these relations put important
constraints on the dynamics, e.g.\ for a background plasma in thermodynamic
equilibrium the pump wave may only decay into waves with lower frequencies.

In the present work three-wave interaction in a homogenous magnetized plasma
is studied using the simplest form of quantum hydrodynamic equations, but
with a slight generalization of the Bohm de Broglie term such that it
depends on a free parameter. The exchange of wave energies among the three
waves are calculated, and the conditions under which the Manley-Rowe
relations are fulfilled is found. The results are compared with previous
works \citep{Murtza-2013,Larsson-1973,Larsson-1977}, and our findings are used to draw general
conclusions regarding the mathematical structure of quantum hydrodynamics.

\section{Quantum hydrodynamics and the Manley-Rowe relations}
\label{chapter2}
The most simple quantum hydrodynamic equations \citep%
{Haas-book,Manfredi,Lundin} reads%
\begin{equation}
\frac{\partial n}{\partial t}+\nabla \cdot \left( n\mathbf{v}\right) =0
\label{cont-1}
\end{equation}%
\begin{equation}
\left( \frac{\partial }{\partial t}+\mathbf{v}\cdot \nabla \right) \mathbf{%
v} = \frac{q}{m} \left( \mathbf{E}+\mathbf{v}\times \mathbf{B}\right) -\frac{\nabla P}{nm}+%
\frac{\hbar ^{2}}{2m^2}\nabla \left( \frac{1}{\sqrt{n}}\nabla ^{2}\sqrt{n}%
\right)  \label{momentum-1}
\end{equation}%
where $n$ is the number density, $\mathbf{v}$ is the fluid velocity, $q$ and 
$m$ are the particle charge and mass, $\mathbf{E}$ and $\mathbf{B}$ are the
electric and magnetic field, $P$ is the pressure and $h=2\pi \hbar $ is
Planck's constant. The last term in Eq. (\ref{momentum-1}) is the Bohm de
Broglie force which normally can be neglected for ions due to the mass
dependence. Eqs. (\ref{cont-1}) and (\ref{momentum-1}) for each species are
complemented by the standard Maxwell equations and an equation of state for
the pressure. An often used simple relation is 
\begin{equation}
\left( \frac{P}{P_{0}}\right) =\left( \frac{n}{n_{0}}\right) ^{\gamma }
\label{pressure-1}
\end{equation}%
which includes isothermal ($\gamma =1$), classical adiabatic ($\gamma =3$)
or Fermi pressures ($\gamma =5/3$) as special cases. Here $P_{0}$ and $n_{0}$
are the unperturbed pressure and number density. Typically when the Bohm de
Broglie force is significant, the thermodynamic temperature $T$ is smaller
than the Fermi temperature $T_{F}=(\hbar ^{2}/2m_{e})(3\pi^{2})^{2/3}n^{2/3}/k_{B}$, 
which makes the Fermi pressure the favored equation of state in quantum plasmas.
While the expression for the Fermi pressure $P_{F} = ( \hbar^{2}/5m_{e}) (3 \pi^{2} )^{2/3} n^{5/3}$ is well established there is still a degree 
of uncertainty regarding the most accurate factor in
the equation of state for a degenerate plasma. The reason is that the for a
weakly collisional system (as is typically appropriate for a plasma), the
system is not in local thermodynamic equilibrium during the compression by
electromagnetic forces, in which case there is no firm bases for any type of
pressure model. Comparions with kinetic theories based on the Wigner
function \citep{Manfredi} can then favor values of $\gamma \neq 5/3$ even for $%
T\ll T_{F}$. We will not be concerned with the best value of $\gamma \,\ $in
the rest of the manuscript, and simply note that for a degenerate plasma we
have $1\lesssim \gamma \lesssim 3$.

Eqs. (\ref{cont-1}) and (\ref{momentum-1}) can be derived from the Schr\"odinger
equation using a Madelung ansatz for the wave function \citep%
{Manfredi,Haas-Manfredi}, where the wave function amplitude become the
square root of the number density and the gradient of the phase is closely
related to the fluid velocity. While the Bohm de Broglie force comes out
straightforwardly from the single particle Schr\"odinger equation, the
derivation of (\ref{momentum-1}) depend on the possibility to interchange
the ordering between averaging over particles and taking spatial derivatives
(see e.g. Eq. (4.30) of Ref. \citep{Manfredi}). While such an interchange
sometimes can be justified, this step becomes questionable when the Bohm de
Broglie force is large, in which case Eq. (\ref{momentum-1}) lacks a firm basis.

Another means to derive quantum hydrodynamics equations is to take moments
of the Wigner function \citep{Manfredi,Gardner,Moment-1,Moment-2}. Such a
procedure can to some extent lend support to Eq. (\ref{momentum-1}), but
depending on technical details it may also generate evolution equations that
deviate considerably from the ones presented here. In particular the quantum
effect may occur firstly in the heat flux equation, not already in the
momentum equation \citep{Moment-1,Moment-2}. A general problem when using
moment expansions is that typically truncation of the series depends on
physical insights rather than mathematical rigor. In the limit of small
collisions the truncation must necessarily involve rather crude
approximations, since the effects of wave-particle interaction (which is
dropped in the fluid limit) is not small in general. In such a scenario when
no rigorous justification from first principles can be made, the credence of
the fluid equations can be determined on two grounds. Firstly, that there is
reasonable agreement with kinetic theory in most situations. Secondly, that
the mathematical structure of the fluid equations is sound. The first
criterion is discussed e.g. in Ref. \citep{Manfredi}, where a good agreement
of (\ref{cont-1}) and (\ref{momentum-1}) with kinetic theory is found for
some model problems. The second criterion is usually deemed to be fulfilled
if proper conservations laws for momentum, energy and angular momentum are
obeyed. Here we would like to extend these requirements on the mathematical
structure, and also demand that the basic equations fulfill the Manley-Rowe
relations \citep{Manley-Rowe} when nonlinear three-wave interaction \citep%
{Weiland-Wilhelmsson,Larsson-1973,Larsson-1977,Brodin-88} is studied.

Let us consider three waves with frequencies and wave numbers $(\omega
_{(i)},\mathbf{k}_{(i)})$ $i=1,2,3$, that propagate in an homogenous
magnetized plasma. We let the frequencies and wave numbers be related
through 
\begin{eqnarray}
\omega _{(3)} &=&\omega _{(1)}+\omega _{(2)}  \label{Freq} \\
\mathbf{k}_{(3)} &=&\mathbf{k}_{(1)}+\mathbf{k}_{(2)}  \label{Wave-vect}
\end{eqnarray}%
which correspond to energy and momentum conservation respectively, in case
we make a quantum mechanical interpretation. The consistency of a quantum
mechanical interpretation depends on the Manley-Rowe relations, however.
According to the Manley-Rowe relations the change of energy (denoted by $W_{(i)}$)
of each wave must be in direct proportion to its
frequency, such that we can imagine wave interaction taking place one quanta
at a time. Thus in terms of the wave energies the Manley-Rowe relations can
be written 
\begin{equation}
\frac{1}{\omega _{(3)}}\frac{dW_{(3)}}{dt}=-\frac{1}{\omega _{(1)}}\frac{%
dW_{(1)}}{dt} 
=
-\frac{1}{\omega _{(2)}}\frac{dW_{(2)}}{dt} 
. \label{MR-rel}
\end{equation}%
All the common classical plasma models lead to coupling coefficients for
three wave interaction that are consistent with the Manley-Rowe relations,
including the Vlasov equation and multifluid equations of the type (\ref%
{cont-1}) and (\ref{momentum-1}) \textit{but without the Bohm de Broglie term%
}, see e.g. Refs. \citep{Larsson-1973,Larsson-1977,Stenflo-1994}.
Furthermore, requiring that (\ref{MR-rel}) is fulfilled can be used as a
means for separating useful plasma models from less physical ones. For a
concrete example, see e.g. Ref. \citep{Brodin-thesis} where a class of
pressure tensor models were investigated, and only the sub-class consistent
with (\ref{MR-rel}) were deemed appropriate. In the section below we will
demonstrate that the fluid equations including the Bohm de Broglie term in
general lead to coupling coefficients that fulfill the Manley-Rowe
relations. It should be stressed that this depend on the detailed
mathematical structure of the quantum force. To emphasize this point we will
consider a slightly generalized Bohm de Broglie term given by 
\begin{equation}
\frac{\hbar ^{2}}{2m}\nabla \left( \frac{1}{n^{\xi }}\nabla ^{2}n^{\xi
}\right) .  \label{general}
\end{equation}%
As we will see below, the Manley-Rowe relations will be fulfilled if and
only if $\xi =1/2$, in which case Eq. (\ref{general}) agrees with the
standard form displayed in Eq. (\ref{momentum-1}) This supports the idea
that fulfillment of the Manley-Rowe relations is a highly useful criterion
in separating physical models from non-physical ones.

\section{The coupled three wave equations}

\subsection{Preliminaries}
In general we consider our variables as given by the sum of a unperturbed background and a small perturbation, e.g. $n=n_0 + \delta n$, where index $0$ denotes the unperturbed value. The background plasma is time-independent and homogenous with zero drift velocities, and the unperturbed magnetic field is $\mathbf{B}_{0}=B_{0}\mathbf{\hat{z}}$. The perturbations consist of contributions from all waves, $\delta n = \sum_{j=1}^{3}n_{(j)}(t)\exp [i(\mathbf{k}_{(j)}\cdot \mathbf{r} -\omega _{(j)}t)]+\mathrm{c.c}.,$ where $\mathrm{c.c.}$ denotes the complex conjugate and the time dependence of the amplitudes are assumed to be slow compared to the wave frequency.
Firstly limiting ourselves to only time-dependent
amplitudes simplifies some of the technical aspects of the derivation. A
generalization to a weak spatial dependence of the amplitudes is easily
included by the substitution $\partial /\partial t\rightarrow \partial
/\partial t+\mathbf{v}_{g(j)}\cdot \nabla $. Here the index $(j)$ on the group
velocity $\mathbf{v}_{g(j)}$ is $j=1,2,3$ depending on which wave amplitude
the derivative is acting.
Next we make an amplitude expansion keeping only up to second order terms. Writing linear quantities on the left hand side, and nonlinear on the right hand side, the momentum equation reads 
\begin{eqnarray}
 && \frac{\partial \delta \mathbf{v} }{\partial t} 
 - \frac{q}{m} (\delta \mathbf{E} + \delta\mathbf{v} \times \mathbf{B}_0) 
 + \frac{\gamma P_0}{n_0^2m } \nabla \delta n 
 - \frac{\xi \hbar^2}{2m^2 n_0} \nabla \nabla^2 \delta n
 \nonumber \\ 
&=& 
 - (\delta \mathbf{v} \cdot \nabla ) \delta \mathbf{v} 
 + \frac{q}{m} \delta \mathbf{v} \times \delta \mathbf{B} 
 - \frac{\gamma (\gamma-2) P_0}{n_0^3m} \delta n \nabla \delta n 
 \nonumber \\ &&
- \frac{ \xi \hbar^2}{2m^2 n_0^2} 
\left( \delta n \nabla \nabla^2 \delta n + \nabla^2 \delta n \nabla \delta n + 2( 1 - \xi)  \left( \nabla \delta n \cdot \nabla \right) \nabla \delta n  \right)
\label{Momentum-2}
\end{eqnarray}

Before we start the nonlinear analysis it is convenient to introduce the
expression for the wave energy densities. These are%
\begin{eqnarray}
W_{(i)}&=&
\frac{\epsilon _{0}}{2}\mathbf{E}_{(i)}\cdot \mathbf{E}_{(i)}^{\ast} 
+ \frac{1}{2\mu _{0}}\mathbf{B}_{(i)}\cdot \mathbf{B}_{(i)}^{\ast }
\nonumber \\ && 
+ \sum_{s} \left[ 
\frac{m_{s}n_{0s}}{2}\mathbf{v}_{(i)s}\cdot \mathbf{v}_{(i)s}^{\ast } + %
\left( \frac{\gamma _{s}P_{0s}}{2n_{0s}^2}+\frac{\xi \hbar ^{2}k_{(i)}^{2}}{%
4m_{s}n_{0s}}\right) n_{(i)s}n_{(i)s}^{\ast }  
\right]
\label{Wave-energy-1}
\end{eqnarray}%
where the star denotes complex conjugate. The expression for the wave energy
densities can be deduced by demanding that $W_{(i)}$ is conserved to all
orders in the slow time derivative (i.e.\ acting on the wave amplitudes) when
the nonlinearities are neglected. 
From the dispersion relation, where the wave
frequency becomes real in the absence of dissipative mechanisms, one can of
course deduce that the different sub-parts of the wave energy are conserved
separately in the absence of nonlinear interactions. However, the wave
energy (\ref{Wave-energy-1}) is the unique expression that can be shown to
be conserved without using the linear dispersion relation. 

Formally all species are treated equivalently in Eqs. (\ref{Momentum-2}) and (\ref%
{Wave-energy-1}). The fact that electrons may be described quantum
mechanically and ion classically can be accounted for in the final result by
choosing $\gamma _{s}$ differently for electrons and ions, and dropping the
Bohm de Broglie term altogether for ions.

\subsection{The Manley Rowe relations}
Including only the linearized terms of the left hand side in (\ref%
{Momentum-2}), as well as in the continuity equation and Maxwell's
equations, the wave energy of each wave is conserved. Including the
quadratically nonlinear terms of the right hand sides, the rate of change of
each wave energy becomes proportional to terms that are cubic in the
amplitude. Only the resonant cubic terms that survives averaging over
several wave periods are kept. Thus the energy change of wave 3 directly
associated with the electric field can be written%
\begin{eqnarray}
&& \frac{\epsilon _{0}}{2} \mathbf{E}_{(3)}^{\ast } \cdot \frac{\partial \mathbf{E}_{(3)}}{\partial t}
	+ \mathrm{c.c}
\nonumber \\ &&
	= \frac{\epsilon _{0}c^{2}}{2}\mathbf{E}_{(3)}^{\ast } 
		\cdot \left[ \nabla \times \mathbf{B}_{(3)}
	- \mu_{0}\sum_{s}q_{s} \left(n_{0s}\mathbf{v}_{(3)s}
	+ n_{(1)s}\mathbf{v}_{(2)s}+n_{(2)s}\mathbf{v}_{(1)s} \right) \right] + \mathrm{c.c}.  
\label{example}
\end{eqnarray}%
in accordance with Eqs.\ (\ref{Freq}) and (\ref{Wave-vect}). As the terms
that are quadratic in the wave fields will cancel when all source terms are
considered, only the cubic terms of the right hand side are of interest
here. Treating the other energy terms in the same manner, we thus find that $%
\mathrm{d}W_{3}/\mathrm{d}t$ becomes proportional to a large number of cubic
terms. Simplifying this expression using linear approximations (e.g.\ $%
n_{(j)s}=n_{0s}\mathbf{k}_{(j)}\cdot \mathbf{v}_{(j)s}/\omega _{(j)}$, etc)
in the cubic terms, we obtain after some lengthy algebra 
\begin{eqnarray}
	\frac{\mathrm{d}W_{(3)}}{\mathrm{d}t}
	&=& 
	\omega _{(3)} \sum_s \bigg[
	-\frac{im_{s}}{2} 
	\Big(
		n_{(1)s}\mathbf{v}_{(2)s}\cdot \mathbf{v}_{(3)s}^{\ast }
		+ n_{(2)s} \mathbf{v}_{(1)s} \cdot \mathbf{v}_{(3)s}^{\ast }
		+ n_{(3)s}^{\ast }\mathbf{v}_{(1)s} \cdot \mathbf{v}_{(2)s}
	\Big)
\nonumber \\ && \quad \quad \quad \quad
	- \frac{i \gamma_{s} (\gamma_{s} - 2) P_{0s} }{n_{0s}^{3}} n_{(1)s} n_{(2)s} n_{(3)s}^{\ast }
\nonumber \\ &&	
	\quad \quad \quad \quad
	+ \frac{i \xi \hbar ^{2}}{8m_{s}n_{0s}^{2}}
	\left[
		k_{(1)}^{2} + k_{(2)}^{2} + k_{(3)}^{2} - ( 2\xi - 1 ) \mathbf{k}_{(1)}
		 \cdot \mathbf{k}_{(2)} 
	\right] n_{(1)s} n_{(2)s} n_{(3)s}^{\ast }  
\nonumber \\&&
\quad \quad \quad \quad
	- \frac{m_{s} \omega _{cs}}{ 2 \omega _{(3)} } n_{0s}
	\left(
		\frac{k_{(2)z}}{ \omega_{(2)} }
		- \frac{k_{(1)z}}{ \omega_{(1)} } 
	\right) \mathbf{v}_{(3)s}^{\ast } \cdot 
	\left(\mathbf{v}_{(1)s}\times \mathbf{v}_{(2)s} \right) \bigg]
	+ \mathrm{c.c}.  
\label{eq:wave_3_power_density_res}
\end{eqnarray}%
Equation (\ref{eq:wave_3_power_density_res}) is our main result, together with
the similar expressions for $dW_{(1,2)}/dt$ that can be obtained directly
from (\ref{eq:wave_3_power_density_res}) using the symmetry between $\omega
_{(1)}$, $\omega _{(2)}$ and $-\omega _{(3)}$ as well as between $\mathbf{k}%
_{(1)}$, $\mathbf{k}_{(2)}$ and $-\mathbf{k}_{(3)}$. When $\hbar \rightarrow
0$ Eq. (\ref{eq:wave_3_power_density_res}) agrees with Ref. \citep%
{Larsson-1977}. Furthermore, the corresponding expression for $W_{(1,2)}$
confirms that the Manley-Rowe relations (\ref{MR-rel}) are fulfilled when $%
\hbar \rightarrow 0$.
At a first glance the last term of (\ref{eq:wave_3_power_density_res}) seems to be in conflict with (\ref{MR-rel}) (i.e. the symmetry between
waves 1, 2 and 3 is not explicit), but simple manipulations using Eqs.\ (\ref{Freq}) and (\ref{Wave-vect})
quickly confirms that the term is in full agreement with (\ref{MR-rel}). 
The quantum term on the other
hand has two very different contributions. The first term (proportional to $%
k_{(1)}^{2}+k_{(2)}^{2}+k_{(3)}^{2}$) is obviously in agreement with (\ref%
{MR-rel}). However, the second term proportional to $(2\xi -1)$ must
vanish for the Manley-Rowe relations to hold, i.e.\ we must have $\xi =1/2$
. Thus we can confirm that fulfillment of (\ref{MR-rel}) can be used as a
criterion for disregarding unphysical models. From now on we limit ourselves
to the standard Bohm de Broglie term with $\xi =1/2$, in which case 
\begin{equation}
\frac{1}{\omega _{(3)}}\frac{dW_{(3)}}{dt}=-\frac{1}{\omega _{(1,2)}}\frac{%
dW_{(1,2)}}{dt}=
V+\mathrm{c.c.}  \label{MR-2}
\end{equation}%
with 
\begin{eqnarray}
V &=& 
\sum_s \bigg[  
-\frac{ i m_{s} }{ 2 }
\left( n_{(1)s } \mathbf{v}_{(2)s} \cdot \mathbf{v}_{(3)s}^{\ast }
+ n_{(2)s} \mathbf{v}_{(1)s} \cdot \mathbf{v}_{(3)s}^{\ast}
+ n_{(3)s}^{\ast }\mathbf{v}_{(1)s} \cdot \mathbf{v}_{(2)s} \right)
\nonumber \\
&&
-\frac{i\gamma _{s}(\gamma _{s}-2)P_{0s}}{n_{0s}^{3}}%
n_{(1)s}n_{(2)s}n_{(3)s}^{\ast }
+\frac{i \hbar ^{2}}{16m_{s}n_{0s}^{2}}
\left[ k_{(1)}^{2}+k_{(2)}^{2}+k_{(3)}^{2} \right] n_{(1)s} n_{(2)s}n_{(3)s}^{\ast }
\nonumber \\ 
&&
- \frac{m_{s}\omega _{cs}}{2\omega _{(3)}}n_{0s}\left( \frac{k_{(2)z}}{\omega
_{(2)}}-\frac{k_{(1)z}}{\omega _{(1)}}\right) \left[ \mathbf{v}_{(3)s}^{\ast
}\cdot \left(\mathbf{v}_{(1)s}\times \mathbf{v}_{(2)s} \right) \right]  \label{Final-V}
\end{eqnarray}%
The property (\ref{MR-2}) has important consequences. Firstly, it means that
a quantum interpretation of the three-wave interaction process is possible,
as noted above. This has the further implication that parametric decay
occurs from higher to lower frequencies, unless the wave energy density is
negative, which can only occur if the background plasma has a free energy
source. Three-wave interaction in homogenous plasmas using quantum
hydrodynamic equations has been considered previously by Ref. \citep%
{Murtza-2013}, specifically focusing on the parametric decay of Langmuir
waves in magnetized plasmas. However, their calculations did not produce
Manley-Rowe symmetric formulas, and thus our above results is an improvement
in this respect. Furthermore, Eq.\ (\ref{Final-V}) cover all types of waves (Alfen waves, Whistler waves, Extra-ordinary, etc.) and thus represents an extensive generalization of previous work.

\subsection{Three wave equations}
In the previous sub-section we showed that the Manley-Rowe relations are
fulfilled for the physical case of $\xi =1/2$. However, in order to do
practical calculations of wave interactions (e.g.\ to find growth rates for
parametric instabilities), we first need to rewrite the equations in terms
of the wave amplitudes rather than wave energy densities. For this purpose
we note that the wave energy densities can be written as $W=\varepsilon
_{0}E_{j}^{\ast }\partial (\omega D_{ij})/\partial \omega )E_{i}$ for each
wave,
where we denote cartesian componets $x,y,z$ with
index $1,2,3$ in order to use the summation convention (a closely related and often used expression for the
wave energy that is equivalent is $W=\varepsilon _{0}(1/\omega )E_{j}^{\ast
}\partial (\omega ^{2}\varepsilon _{ij})/\partial \omega )E_{i}$, where $%
\varepsilon _{ij}$ is the dielectric tensor). 
The electric
field eigenvectors fulfill $D_{ij}E_{j}=0$ with 
\begin{equation}
D_{ij}=\left( 1-\frac{k^{2}c^{2}}{\omega ^{2}}\right) \xi _{ij}+\frac{%
k_{i}k_{j}c^{2}}{\omega ^{2}}+\sum_{s}\chi _{ij}
\end{equation}%
and the susceptibility tensor $\chi _{ij}$ for each species is given
by 
\begin{eqnarray}
&&\mathbf{\chi }=-\frac{\omega _{p}^{2}}{\Omega _{s}^{4}} \times 
\nonumber \\ 
&&
\left[ \! \!
\begin{array}{ccc}
\omega^{2} - (k^{2} - k_{x}^{2} )V_{s}^{2} 
& 
k_{x}k_{y}V_{s}^{2} + \frac{i\omega_{cs}}{\omega }(\omega^{2}-k_{z}^{2}V_{s}^{2}) 
& k_{x}k_{z}V_{s}^{2}+\frac{i\omega _{cs}}{\omega }k_{y}k_{z}V_{s}^{2} \\ 
&  &  \\ 
k_{x}k_{y}V_{s}^{2}-\frac{i\omega _{cs}}{\omega }(\omega
^{2}-k_{z}^{2}V_{s}^{2}) & \omega ^{2}-(k^{2}-k_{y}^{2})V_{s}^{2} & 
k_{y}k_{z}V_{s}^{2}-\frac{i\omega _{cs}}{\omega }k_{x}k_{z}V_{s}^{2} \\ 
&  &  \\ 
k_{x}k_{z}V_{s}^{2}-\frac{i\omega _{cs}}{\omega }k_{y}k_{z}V_{s}^{2} & 
k_{y}k_{z}V_{s}^{2}+\frac{i\omega _{cs}}{\omega }k_{x}k_{z}V_{s}^{2} & \omega
^{2}-\omega _{cs}^{2}-V_{s}^{2}(k^{2}-k_{z}^{2})%
\end{array}%
\! \!
\right],
\nonumber \\ &&
\end{eqnarray}%
where we have defined 
\begin{eqnarray}
\Omega _{s}^{4} &=&\omega ^{2}(\omega ^{2}-V_{s}^{2}k^{2})-\omega
_{cs}^{2}(\omega ^{2}-V_{s}^{2}k_{z}^{2}) \\
V_{s}^{2} &=& \frac{\gamma P_{0s}}{m_s n_{0s}}+\frac{\hbar ^{2}k^{2}}{4m_{s}^{2}}.
\end{eqnarray}
The linear susceptibility in a fluid theory
including the Bohm de Broglie force has been computed in Ref. \citep{Lundin}.
Here we have generalized this expression to arbitrary cartesian coordinate
axes, since we cannot chose a coordinate axis along the perpendicular
wavenumber for mote than one of the interacting waves in general.
Finally the dispersion relation for each wave is determined by 
\begin{equation}
D(\omega ,\mathbf{k})\equiv \det D_{ij}=0.  \label{E-polarization}
\end{equation}%
Now we want to express all quantities appearing in (\ref{MR-2}) and (\ref%
{Final-V}) in terms of a single variable representing the wave amplitude of
each wave. Somewhat arbitrarily we can pick the z-component of the electric
fields, but the procedure outlined below works for any component of the
electric field. Firstly using $D_{ij}E_{j}=0$ we can express $E_{x}$ and $%
E_{y}$ in terms of $E_{z}$ for each wave. Together with $v_{i}=-i \omega \epsilon_0 \chi
_{ij}E_{j} / qn_0 $ this gives all velocity components in
terms of $E_{z}$, and the density perturbation is obtained in terms of $%
\delta n=n_0 k_{i}v_{i}/\omega $. The remaining quantity needed is the wave energy
density, which with the help of $W=\varepsilon _{0}E_{j}^{\ast }\partial
(\omega D_{ij})/\partial \omega )E_{i}$ is written as
\begin{equation}
W_{(3)}=\frac{\varepsilon _{0} \omega_3}{(D_{xx}D_{yy}-D_{xy} D_{yx})}\frac{%
\partial D(\omega _{3},\mathbf{k}_{3}\mathbf{)}}{\partial \omega _{3}}%
E_{(3)z}E_{(3)z}^{\ast }  \label{W-3}
\end{equation}
for wave 3. As a consequence all variables appearing in (\ref{MR-2}) and (%
\ref{Final-V}) can be expressed in terms of the z-component of the electric
field amplitudes, in which case Eqs. (\ref{MR-2}) and (\ref{Final-V}) can be
rewritten as%
\begin{eqnarray}
\frac{dE_{(1,2)z}}{dt} &=&\alpha _{(1,2)}E_{(3)z}E_{(2,1)z}^{\ast } \\
\frac{dE_{(3)z}}{dt} &=&\alpha _{(3)}E_{(1)z}E_{(2)z} , 
\end{eqnarray}%
where we now allow for spatially dependent amplitudes such that 
\begin{equation}
	\frac{ dE_{(j)z} }{dt} = \left( \frac{\partial}{\partial t} 
	+ \mathbf{v}_{g(j) } \cdot \nabla \right) E_{(j)z} . 
\end{equation} 
It is straightforward to find the general expressions for
the coupling coefficients $\alpha _{(1,2,3)}$from formulas (\ref%
{MR-2}), (\ref{Final-V}) and (\ref{W-3}) and the procedure outlined above.
However, in order to obtain comparatively simple and illustrative formulas,
we consider the special case where the plasma is unmagnetized, $B_{0}=0$,
and waves 1 and 3 are Langmuir waves. Furthermore we let the plasma be degenerate, i.e. for a 3D Fermi gas $P_0=n_0 m v_F^2/5$ and $\gamma=5/3$, where we have used the thermodynamic equilibrium pressure (see discussion in Section \ref{chapter2}).
In this case the general dispersion
relation (\ref{E-polarization}) reduces to%
\begin{equation}
\omega _{(1,3)}^{2}=\omega _{p}^{2}+\frac{1}{3}k_{(1,3)}^{2}v_{F}^{2}+\frac{\hbar
^{2}k_{(1,3)}^{4}}{4m^{2}}  \label{DR-Langmuir}
\end{equation}%
when the corrections due to the ion \ motion is neglected. Wave 2 is a
low-frequency ion-acoustic wave fulfilling the approximate dispersion
relation 
\begin{equation}
\omega _{(2)}^{2}=\frac{\omega _{pi}^{2}}{1+\omega
_{pe}^{2}/(\frac{1}{3}k_{(2)}^{2}v_{F}^{2}+\hbar ^{2}k_{(2)}^{4}/4m_{e}^{2})}
\label{DR-ion-ac}
\end{equation}%
where we have set the ion temperature to zero and let $m_{e}/m_{i}%
\rightarrow 0$, but avoided the approximation of quasi-neutrality in order
to allow for short wavelengths. Making the corresponding approximations in (%
\ref{Final-V}) and (\ref{W-3}) we obtain the coupled equations:%
\begin{eqnarray}
	\frac{ \partial \phi _{(3)} }{ \partial t } &=& 
	- \frac{i q_{e}  }{2 m_{e} \omega _{(3)} k_{(3)}^{2} \omega _{pe}^{2} }
	\tilde V \phi _{(1)} \phi _{(2)} 
\\
\frac{\partial \phi _{(2)}}{\partial t}&=&
\frac{ i q_{e}\omega _{(2)}^{3} }{2 m_{e} k_{(2)}^{2} \omega _{pi}^{2} \omega _{pe}^{4} }
\tilde V \phi _{(1)}^{\ast }\phi _{(3)} 
\\
\frac{\partial \phi _{(1)}}{\partial t} &=& 
\frac{i q_{e}}{2 m_{e}\omega _{(1)} k_{(1)}^{2} \omega _{pe}^{2}}
\tilde V \phi _{(3)}\phi _{(2)}^{\ast }, 
\end{eqnarray}
where 
\begin{equation} 
	\tilde V = \left( 1 - \frac{ \omega _{pi}^{2} }{ \omega _{(2)}^{2} } \right) 
	\left( \omega _{(1)} \omega _{(3)}  k_{(2)}^{2} \mathbf{k}_{(1)} \cdot \mathbf{k}_{(3)}
	- \left[ \frac{v_F^2}{9} 
	+ \frac{ \hbar^{2} }{ 8m_{e}^{2} }( k_{(1)}^{2} + k_{(2)}^{2} + k_{(3)}^{2} ) \right] k_{(1)}^{2} 		k_{(2)}^{2} k_{(3)}^{2} \right) 
\end{equation}

As usual these equations can be used to compute growth rates
and threshold values for parametric instabilities (if the pump wave has a
finite width or a damping mechanism due to e.g. collisions is added), see
e.g. Ref. \citep{Weiland-Wilhelmsson,Brodin-Scripta-88}.

\section{Concluding discussion}
In this paper we have focused on the Manley-Rowe relations in quantum
hydrodynamics. Our starting point has been that basic equations that are
physically sound should produce coupling coefficients for three-wave interaction that
obey these relations. As discussed by e.g.\ Ref.\ \citep{Larsson-Hamiltonian}
fulfillment of the Manley-Rowe relations comes from an underlying
Hamiltonian structure. For classical plasmas, it is illustrated rather
clearly in Ref.\ \citep{Stenflo-1994} that the Manley-Rowe relations are
satisfied for arbitrary wave propagation in hot magnetized plasmas with a
uniform background. Moreover the Manley-Rowe relations are more general than
expected, i.e. they are sometimes applicable outside their expected range of
validity, see Ref.\ \citep{Kaufman-1975}. Generalized Manley-Rowe relations
are also valid fo non-uniform plasmas \citep%
{Kaufman-1979,Lindgren-1981,Aliev-1990} and somewhat surprisingly also for
turbulent plasma (see the rather instructive paper by Ref.\ \citep%
{Vladimirov-1997}).

Nevertheless, the derivation of standard quantum hydrodynamic equations
contain steps that can be questioned when both the pressure and particle
dispersive effects are large. Hence, it is not obvious that such equations
preserves a physically sound structure, i.e. obeys the Manley-Rowe
relations. As demonstrated by Eqs.\ (\ref{MR-2}-\ref{Final-V}), however,
these relations are indeed fulfilled when the standard Bohm de Broglie term
is used to describe particle dispersive effects. This is \textit{not the case%
}, however, in case the Bohm de Broglie term is replaced by a slightly
generalized expression, which demonstrates that fulfillment of the
Manley-Rowe relations is a useful criterion in separating acceptable models
from unphysical ones. Besides these theoretical aspects we note that our resutls extends previous works on three-wave interaction based on classical fluid equations \citep{Larsson-1977} to cover quantum hydrodynamics. The quantum contribution to the coupling coefficient
is important for short wavelengths (of the order of the thermal de Broglie
wavelength), and for a quantum parameter $H=\hbar \omega _{p}/k_{B}T$ that
is not much smaller than unity. See e.g.\ Refs.\ \citep%
{Haas-book,Manfredi,Shukla-Eliasson,Shukla-Eliasson-RMP} for a thorough
discussion of systems that fit this description.



\bibliographystyle{plain}

\bibliography{2013-12-16Submissiontoarxiv}

\begin{thebibliography}{10}

\bibitem{Aliev-1990}
Yu.~M. Aliev and G.~Brodin.
\newblock Instability of a strongly inhomogeneous plasma.
\newblock {\em Phys. Rev. A}, 42:2374--2378, Aug 1990.

\bibitem{Atwater-Plasmonics}
H.~A. Atwater.
\newblock The promise of plasmonics.
\newblock {\em Scientific American}, 296:56--62, 2007.

\bibitem{Brodin-Scripta-88}
G.~Brodin and L.~Stenflo.
\newblock Parametric instabilities of finite amplitude alfv{\'e}n waves.
\newblock {\em Physica scripta}, 37(1):89, 1988.

\bibitem{Brodin-88}
G.~Brodin and L.~Stenflo.
\newblock Three-wave coupling coefficients for mhd plasmas.
\newblock {\em Journal of plasma physics}, 39(02):277--284, 1988.

\bibitem{Brodin-thesis}
G.~Brodin and L.~Stenflo.
\newblock Three-wave coupling coefficients for magnetized plasmas with pressure
  anisotropy.
\newblock {\em J. Plasma Phys}, 41:199--208, 1989.

\bibitem{Dysthe-1977}
K.~B. Dysthe, E.~Leer, J.~Trulsen, and L.~Stenflo.
\newblock Stimulated brillouin scattering in the ionosphere.
\newblock {\em Journal of Geophysical Research}, 82(4):717--718, 1977.

\bibitem{Gardner}
C.~L. Gardner.
\newblock The quantum hydrodynamic model for semiconductor devices.
\newblock {\em SIAM Journal on Applied Mathematics}, 54(2):409--427, 1994.

\bibitem{glenzer-redmer}
Siegfried~H. Glenzer and Ronald Redmer.
\newblock X-ray thomson scattering in high energy density plasmas.
\newblock {\em Rev. Mod. Phys.}, 81:1625--1663, Dec 2009.

\bibitem{Haas-book}
F.~Haas.
\newblock {\em Quantum Plasmas}.
\newblock Springer, New York, 2011.

\bibitem{Moment-1}
F.~Haas, M.~Marklund, G.~Brodin, and J.~Zamanian.
\newblock Fluid moment hierarchy equations derived from quantum kinetic theory.
\newblock {\em Physics Letters A}, 374(3):481--484, 2010.

\bibitem{Moment-2}
F.~Haas, J.~Zamanian, M.~Marklund, and G.~Brodin.
\newblock Fluid moment hierarchy equations derived from gauge invariant quantum
  kinetic theory.
\newblock {\em New Journal of Physics}, 12(7):073027, 2010.

\bibitem{Astrophysics2}
A.~K. Harding and D.~Lai.
\newblock Physics of strongly magnetized neutron stars.
\newblock {\em Reports on Progress in Physics}, 69(9):2631, 2006.

\bibitem{Kadomtsev}
B.~B. Kadomtsev.
\newblock {\em Plasma turbulence}.
\newblock Academic Press, London, 1965.

\bibitem{Kaufman-1975}
A.~N. Kaufman and L.~Stenflo.
\newblock Action conservation in the presence of a high-frequency field.
\newblock {\em Plasma Physics}, 17(5):403, 1975.

\bibitem{Kaufman-1979}
A.~N. Kaufman and L.~Stenflo.
\newblock Wave coupling in cold nonuniform magnetoplasma.
\newblock {\em Physica Scripta}, 19:523, 1979.

\bibitem{Astrophysics}
C.~Kouveliotou, S.~Dieters, T.~Strohmayer, J.~Van~Paradijs, G.~J. Fishman,
  C.~A. Meegan, K.~Hurley, J.~Kommers, I.~Smith, D.~Frail, et~al.
\newblock An x-ray pulsar with a superstrong magnetic field in the soft
  $\gamma$-ray repeater sgr1806- 20.
\newblock {\em Nature}, 393(6682):235--237, 1998.

\bibitem{Kruer-book-laser-plasma}
W.~L. Kruer.
\newblock {\em The physics of laser plasma interactions}.
\newblock Addison-Wesley, 1988.

\bibitem{Larsson-Hamiltonian}
J.~Larsson.
\newblock A new hamiltonian formulation for fluids and plasmas. part 3.
  multifluid electrodynamics.
\newblock {\em Journal of Plasma Physics}, 55:279--300, 4 1996.

\bibitem{Larsson-1973}
J.~Larsson and L.~Stenflo.
\newblock Three-wave interactions in magnetized plasmas.
\newblock {\em Beitr{\"a}ge aus der Plasmaphysik}, 13(3):169--181, 1973.

\bibitem{Lashmore-Davies-book-chapter}
C.N. Lashmore-Davies.
\newblock 14 - nonlinear laser plasma interaction theory.
\newblock In R.~D. Gill, editor, {\em Plasma Physics and Nuclear Fusion
  Research}, pages 319 -- 354. Academic Press, London, 1981.

\bibitem{Lindgren-1981}
T.~Lindgren, J.~Larsson, and L.~Stenflo.
\newblock Three-wave coupling coefficients for non-uniform plasmas.
\newblock {\em Journal of Plasma Physics}, 26:407--418, 1981.

\bibitem{Lundin}
J.~Lundin, J.~Zamanian, M.~Marklund, and G.~Brodin.
\newblock Short wavelength electromagnetic propagation in magnetized quantum
  plasmas.
\newblock {\em Physics of plasmas}, 14:062112, 2007.

\bibitem{Manfredi}
G.~Manfredi.
\newblock How to model quantum plasmas.
\newblock In T.~Passot, C.~Sulem, and P.-L. Sulem, editors, {\em Topics in
  Kinetic Theory}. Fields Institute Communications, 2005.

\bibitem{Haas-Manfredi}
G.~Manfredi and F.~Haas.
\newblock Self-consistent fluid model for a quantum electron gas.
\newblock {\em Physical Review B}, 64(7):075316, 2001.

\bibitem{Manfredi-quantum-well}
G.~Manfredi and P.~A Hervieux.
\newblock Autoresonant control of the many-electron dynamics in nonparabolic
  quantum wells.
\newblock {\em Applied Physics Letters}, 91(6):061108--061108--3, 2007.

\bibitem{Manley-Rowe}
J.~M. Manley and H.~E. Rowe.
\newblock Some general properties of nonlinear elements-part i. general energy
  relations.
\newblock {\em Proceedings of the IRE}, 44(7):904--913, 1956.

\bibitem{Mironov-1990}
V.~A. Mironov, A.~M. Sergeev, E.~V. Vanin, and G.~Brodin.
\newblock Localized nonlinear wave structures in the nonlinear photon
  accelerator.
\newblock {\em Phys. Rev. A}, 42:4862--4866, Oct 1990.

\bibitem{Astrophysics1}
D.~M. Palmer, S.~Barthelmy, N.~Gehrels, R.~M. Kippen, T.~Cayton,
  C.~Kouveliotou, D.~Eichler, R.~A. M.~J. Wijers, P.~M. Woods, J.~Granot, Y.~E.
  Lyubarsky, E.~Ramirez-Ruiz, L.~Barbier, M.~Chester, J.~Cummings, E.~E.
  Fenimore, M.~H. Finger, B.~M. Gaensler, D.~Hullinger, C.~B. Krimm,
  H.~Markwardt, J.~A. Nousek, A.~Parsons, S.~Patel, T.~Sakamoto, G.~Sato,
  M.~Suzuki, and J.~Tueller.
\newblock A giant big gamma-ray flare from the magnetar sgr 1806-20.
\newblock {\em Nature}, 434:1107--1109, 2005.

\bibitem{Sagdeev-64}
R.~Z. Sagdeev and A.~Galeev.
\newblock {\em Nonlinear plasma Theory}.
\newblock W. A. Benjamin, New York, 1969.

\bibitem{Murtza-2013}
M.~Shahid, A.~Hussain, and G.~Murtaza.
\newblock A comparison of parametric decay of oblique langmuir wave in high and
  low density magneto-plasmas.
\newblock {\em Physics of Plasmas}, 20:092121, 2013.

\bibitem{Shukla-Eliasson}
P.~K. Shukla and B.~Eliasson.
\newblock Nonlinear aspects of quantum plasma physics.
\newblock {\em Physics-Uspekhi}, 53(1):51, 2010.

\bibitem{Shukla-Eliasson-RMP}
P.~K. Shukla and B.~Eliasson.
\newblock \textit{Colloquium} : Nonlinear collective interactions in quantum
  plasmas with degenerate electron fluids.
\newblock {\em Rev. Mod. Phys.}, 83:885--906, Sep 2011.

\bibitem{Sjolund-67}
A.~Sj\"{o}und and L.~Stenflo.
\newblock Non-linear coupling in a magnetized plasma.
\newblock {\em Zeitschrift f\"{u}r Physik}, 204(3):211--214, 1967.

\bibitem{Stenflo-1994}
L.~Stenflo.
\newblock Resonant three-wave interactions in plasmas.
\newblock {\em Physica Scripta}, 1994(T50):15, 1994.

\bibitem{Stenflo-2004}
L.~Stenflo.
\newblock Comments on stimulated electromagnetic emissions in the ionospheric
  plasma.
\newblock {\em Physica Scripta}, 2004(T107):262, 2004.

\bibitem{Larsson-1977}
L.~Stenflo and J.~Larsson.
\newblock Three-wave coupling coefficients for magnetized plasmas.
\newblock In H.~Wilhelmsson, editor, {\em Plasma physics: nonlinear theory and
  experiments}, volume~36 of {\em Proceedings of Nobel Symposium}, pages
  152--158, New York, 1977. Plenum Press.

\bibitem{Tsytovich}
V.~N. Tsytovich.
\newblock {\em Nonlinear Effects in Plasmas}.
\newblock Plenum Press, New York, 1970.

\bibitem{Vladimirov-1997}
S.~V. Vladimirov and L.~Stenflo.
\newblock Three-wave processes in a turbulent nonstationary plasma.
\newblock {\em Physics of Plasmas}, 4:1249, 1997.

\bibitem{Weiland-Wilhelmsson}
J.~Weiland and H.~Wilhelmsson.
\newblock {\em Coherent non-linear interaction of waves in plasmas}.
\newblock Pergamon Press, New York, 1977.

\bibitem{Spintronics}
S.~A. Wolf, D.~D. Awschalom, R.~A. Buhrman, J.~M. Daughton, S.~von Moln, M.~L.
  Roukes, A.~Y. Chtchelkanova, and D.~M. Treger.
\newblock Spintronics: A spin-based electronics vision for the future.
\newblock {\em Science}, 294(5546):1488--1495, 2001.

\end{thebibliography}

\end{document}